\documentstyle[11pt]{article}
\font\sm=cmr9
\newcommand {\be}{\begin{equation}} 
\newcommand{\fe}{\end{equation}}
\newcommand{\eqn}{\label}

\begin{document}

\font\sm=cmr8

\title{CANCELLATION OF LINEARISED AXION-DILATON
SELF ACTION DIVERGENCE IN STRINGS}

\author{ {\bf Brandon Carter}
\\ D.A.R.C., (UPR 176, CNRS),
\\ Observatoire de Paris, 92 Meudon, France.}

\date{\it Contrib. to 1998 Peyresq meeting; to be published in Int. J. Theor.
Phys.}

\maketitle

{\bf Abstract:}

The force densities exerted on a localised material system by
linearised interaction with fields of axionic and dilatonic type are
shown to be describable very generally by relatively simple expressions
that are well behaved for fields of purely external origin, but that
will be subject to ultraviolet divergences requiring regularisation for
fields arising from self interaction in submanifold  supported
``brane'' type systems. In the particular case of 2-dimensionally
supported, i.e. string type, system in an ordinary 4-dimensional
background it is shown how the result of this regularisation is
expressible in terms of the worldsheet curvature vector $K^\mu$, and
more particularly that (contrary to what was suggested by early work on
this subject) for a string of Nambu Goto type the divergent
contribution from the dilatonic self action will always be directed
oppositely to its axionic counterpart. This makes it possible for the
dilatonic and axionic divergences to entirely cancel each other out (so
that there is no need of a renormalisation to get rid of
``infinities'') when the relevant coupling coefficents are related by
the appropriate proportionality condition provided by the low energy
limit of superstring theory.

\section{Introduction }

This article is intended as a contribution to the clarification of a 
question raised by Dabholkar and Harvey~\cite{DH89} who pointed out that
the finiteness of superstring theory implied a similar finiteness
for the corresponding low energy classical limit theory as constructed 
in terms of Nambu Goto type strings interacting via coupling to
gravitational, dilatonic, and axionic type fields.  What this means
is that  although the gravitational, dilatonic and axionic self
interaction contributions will each be separately divergent in generic 
classical string models, their net effect should cancel out --
so that no ``infinite'' renormalisation is required -- in the special case
of the particular model obtained from the low energy limit of superstring
theory. Although there has always been a general concensus to the effect
that this noteworthy prediction is indeed correct, the question of how the
expected cancellation actually comes about has been the subject of
considerable discord.

Relying on  the straightforward -- but with hindsight obviously flawed --
analysis of Dabholkar and Harvey themselves, the opinion that was most widely
held until recently is succinctly summarised by the assertion~\cite{DGHR90}
that ``Since the quantum answer is zero the classical answer must be zero.
This is indeed true. There are three divergent contributions to the classical
self energy. The dilaton contribution is the same as the axion but the
gravitational contribution is negative and twice as large.'' The flaw in this
particular analysis is that it takes account only of the external
field energy contributions, while neglecting the internal (world sheet
supported) contributions, either because the authors assumed (wrongly) that
they would be relatively negligible, or else because they simply forgot
about them, perhaps due to having started by working out the axion part, 
for which the previous literature was already well developped~\cite{VV87}
and for which it just so happens that there is no internal contribution.   

There was however a noteworthy dissenting opinion: while agreeing that
that total was indeed zero, it was claimed on the basis of a rather
obscure calculation by Copeland, Haws and Hindmarsh~\cite{CHH90} that
-- far from being the same as that of the axion -- the dilaton
contribution cancelled out all by itself, leaving the axion
contribution to be cancelled just by the gravitational contribution. It
appears with hindsight that this nonconformist conclusion was not quite
so wide of the mark as that of Dabholkar and Harvey, but it does not
seem to have been taken so seriously, presumably because of the failure
to make clear the logical reasonning on which it was supposed to have
been based.

It has now become evident that both of these competing alternatives were
incorrect, essentially for the same underlying reason, which was the use of
naive force or energy formulae based on the omission of various
important terms that were wrongly assumed to be negligible or simply
forgotten about.  These overhasty omissions were originally motivated
to a large extent by the technical difficulties involved in actually
evaluating the terms in question, but such difficulties have recently
been alleviated by the introduction of more efficient geometrical
methods. Following the derivation by these methods of the correct
formula for the complete gravitational force contribution~\cite{BC95}
acting on a general string model, it has recently been
found~\cite{CB98} that the divergent part of the gravitational self
force cancels itself out all by itself in the case of a Nambu-Goto
string in 4-dimensions. The implication of this is that -- in order for
the total to vanish in the low energy classical limit derived from
superstring theory -- the corresponding dilaton contribution should
neither be equal to the axion contribution (as asserted by Dabholkar
and Harvey) nor even zero (as asserted by Copeland, Haws and
Hindmarsh)  but in fact exactly opposite to the axion contribution, at
least in a 4-dimensional background.

The purpose of the present work is to provide a direct verification of
this corollary -- i.e. of the mutual cancellation of the relevant
axionic and dilatonic string self action divergences for the linearised
classical limit derived from superstring theory in 4-dimensions.  A
less direct confirmation, and and a generalisation to higher dimensions
(where the total still vanishes, but not the gravitational part on its
own) has already been provided using a new approach based on the use of
an effective action by Buonanno and Damour~\cite{BD98a}. It is
worthwhile to provide further confirmation because the pertinance of
this kind of approach was explicitly -- but unjustly -- cast into doubt
by Copeland, Haws and Hindmarsh~\cite{CHH90} who alleged that ``in
general it is not correct'' to deduce the divergent part of the self
force from the divergent part of the effective action, the reason for
their scepticism being the discrepancies that arose in their own rather
incoherent approach.

The verdict of the present analysis is that use of the effective action
approach is inherently correct after all, and that there are no
discrepancies, provided all the calculations are carried out properly
without omission of relevant terms. The consistency of an approach
based on a force analysis with an approach using an effective action had
previously been demonstrated for electromagnetic interactions in
strings~\cite{C97}. Moreover the the results of the effective action
approach used by Buonanno and Damour~\cite{BD98a} were known from the
outset to be consistent with the detailed force analysis -- as
correctly carried out with all relevant terms included -- for purely
gravitational interactions~\cite{CB98}. In this gravitational case the
detailed relationship between the two kinds of approach has since been
demonstrated explicitly \cite{C98a,C98b}. The present work provides an
analogously explicit demonstration of consistency between the effective
action approach used by Buonanno and Damour~\cite{BD98a} and a detailed
force analysis for the technically simpler cases of dilatonic and
axionic interactions.

\section{Lagrangian for dilatonic and axial current couplings.}

Before considering the divergences that arise from self interaction it is
first necessary to consider the effect of linear interactions with generic fields of 
the kinds with which we are concerned, namely a dilatonic (scalar) field $\phi$
and an axionic (pseudoscalar) field represented by a antisymmetric
Kalb-Ramond type 2-form $B_{\mu\nu}$ say. We shall ultimately be concerned with
the application to the classical low energy limit of superstring theory which
also involves a symmetric gravitational perturbation field $h_{\mu\nu}$.
However due to the assumption of linearity (which will be physically justified
when the fields are sufficiently weak) each of the three pertinant
(dilatonic, axionic, and gravitational) parts can be treated independently 
of the other two. Since the relevant analysis of the gravitational
part is already available~\cite{BC95,CB98,C98a,C98b} the present article will
be restricted to the corresponding analysis for the technically simple
axionic and dilatonic parts. Unlike the dilatonic part, the axionic part
seems to have been correctly treated in the earlier work~\cite{DH89,CHH90},
but we shall work through it again here in order to demonstrate the use
of the neater, more efficient, and therefore less error prone mathematical
formalism that have since been developed~\cite{C96,C97a} and that are
indispensible for avoiding unnecessarily heavy algebra in more complicated
applications such  the gravitational part, and that are helpful even for 
relatively simple applications such as to the dilatonic part dealt with here.

The action governing the kind of system to be analysed here will
consist of a total, ${\cal I}_{_{\rm to}}={\cal I}_{_{\rm ra}}
+{\cal I}_{_{\rm 
ma}}$ in which the first term is a a free radiation contribution
of the form 
\be
{\cal I}_{_{\rm ra}}=\int \hat{\cal L}_{_{\rm ra}}
\Vert g\Vert^{1/2}\, d^4 x \, ,\eqn{1} \fe 
where $\Vert g\Vert$ is the modulus of the divergence of the 4-dimensional
spacetime background metric $g_{\mu\nu}$ as expressed with respect to local
coordinates $x^\mu$, in which the Lagrangian density scalar $\hat{\cal
L}_{_{\rm ra}}$ depends only, in a homogeneous quadratic manner, on the
relevant linearised long range field variables, namely $\phi$ and $B_{\mu\nu}$
in the present instance. The other part of the action is a material
contribution
\be {\cal I}_{_{\rm ma}}=\int \hat{\cal L}_{_{\rm ma}}\Vert g\Vert^{1/2}
\, d^4 x \, ,\eqn{2}\fe 
in which the Lagrangian density scalar $\hat{\cal L}_{_{\rm ma}}$ is
restricted to have a purely linear dependence on these long range field
variables, while also having a generically non-linear dependence on whatever
other variables may be needed to characterise the localised  material system
under consideration, which in the application that follows will be taken to be
string. It will be postulated that the material system is unpolarised in the
technical sense that its dependence on the linearised long range field
variables does not involve derivatives, which means that its Lagrangian
density scalar will have the form
\be \hat{\cal L}_{_{\rm ma}}=\hat{\cal L}+\hat{\cal L}_{_{\rm co}}
\eqn{3a}\fe
in which the primary contribution ${\cal L}$ is entirely independent of the
linearised long range field variables $\phi$ and $B_{\mu\nu}$ while the
coupling term  $\hat{\cal L}_{_{\rm co}}$ will have the homogeneous form
\be \hat{\cal L}_{_{\rm co}}={_1\over^2}\hat W^{\mu\nu}B_{\mu\nu}
+\hat T\phi\, ,\eqn{3}\fe
in which the antisymmetric tensor field $\hat W^{\mu\nu}$ and the scalar field
$\hat T$ are each independent of both $B_{\mu\nu}$ and $\phi$.

Unlike the linear coupling term the homogeneously quadratic free radiation
contribution  ${\cal L}_{_{\rm ra}}$ will involve field 
gradients. This contribution will be given in terms of
covariant differentiation with respect to the background metric
$g_{\mu\nu}$, for which the usual symbol $\nabla$ will be used,  by an
expression of the form 
\be {\cal L}_{_{\rm ra}}=-{1\over
8\pi}\Big(m_{_{\rm D}}^{\, 2}\phi^{;\mu} \phi_{;\mu}+ {1\over 6
m_{_{\rm J}}^{\,2}}J^{\mu\nu\rho}J_{\mu\nu\rho}\Big)\, ,\eqn{4}\fe 
using the notation 
\be \phi_{;\mu}=\nabla_{\!\mu}\phi\, \hskip 1 cm
J_{\mu\nu\rho}=3\nabla_{\![\mu}B_{\nu\rho]}\, ,\eqn{5}\fe 
where square brackets indicate index antisymetrisation. The parameters
$m_{_{\rm D}}$ and $m_{_{\rm J}}$ in this expression are fixed coupling
constants having the dimensions of mass on the understanding that we
are using units in which the speed of light $c$ and the Dirac-Planck
constant $\hbar$ are set to unity.  The quantity $m_{_{\rm D}}$ is what
may conveniently be referred to as the Dicke mass. This dilation mass
scale should not be confused with the dilaton mass, $m_{_\phi}$ say,
which is usually assumed to be very small, and which is simply taken to
be zero in the present work. The Dicke dilation mass scale  $m_{_{\rm
D}}$ is usually supposed to be very large, at least comparable with the
Planck mass defined by  $m_{_{\rm P}}= \hbox{\sm G}^{-1/2}$ where 
$\hbox{\sm G}$ is Newton's
constant.  In particular, if the theory is to be applied to the modern
solar system then there are severe observational limits~\cite{D64} that
can be interpreted as implying that the relevant value of the
dimensionless Brans-Dicke parameter $\omega=2\hbox{\sm G}m_{_{\rm D}}^{\,2}+3/2$
should be very large compared with unity and hence that $m_{_{\rm D}}$
is large compared with $m_{_{\rm P}}$.  

The other mass scale, $m_{_{\rm J}}$, is what may conveniently be referred to
as the Joukowski mass, since the corresponding Kalb-Ramond coupling -- of the
axial kind associated in the context of superstring theory with the names of
Wess and Zumino -- gives rise to a lift force of the type that has long been
well known in the context of aerofoil theory, where it was originally derived
as a corollary of the magnus effect by the Russian theoretician Joukowski.
This mass scale can also usually be supposed to be very large, unlike the
associated axion mass, $m_{_{\rm a}}$ say, with which it should not be
confused. Like the dilaton mass $m_{_\phi}$, the axion mass $m_{_{\rm a}}$ is
usually assumed to be very small, and will be taken to be zero for the
purposes of the present work. The axial current model obtained in this
way is interpretable as the ``stiff'' (Zel'dovich type) limit
(characterised by sound perturbation propagation at the speed of light)
within the more general category of ordinary perfect fluid models.
The Kalb-Ramond Wess-Zumino Joukowski coupling bivector field 
$\hat W^{\mu\nu}$ is interpretable~\cite{C94b} as a vorticity
flux, and must be such as to satisfy a flux conservation law of the form
\be \nabla_{\!\mu}\hat W{^{\mu\nu}}=0\, ,\eqn{6} \fe
in order to ensure invariance under local Kalb-Ramond gauge 
transformations of the form
$B_{\mu\nu}\mapsto B_{\mu\nu}+2\nabla_{\![\mu}\varphi_{\nu]}$
for an arbitrary covector field $\varphi_\nu$. In typical applications
to continuous media~\cite{C94b} this condition is fullfilled by
by a specification of the form $\hat W^{\mu\nu}=\varepsilon^{\mu\nu\rho\sigma}
\chi^+_{\,;\rho}\chi^-_{\,;\sigma}$ where the scalars
$\chi^+$ and $\chi^-$ are two of the intrinsic field variables 
characterising the material system and $\varepsilon^{\mu\nu\rho\sigma}$
is the usual antisymmetric measure tensor induced on the background space
(modulo a choice of signature) by the metric $g_{\mu\nu}$. In such a case,
as in the case of the string type systems for which the relevant formula
for $\hat W^{\mu\nu}$ will be described below, the corresponding coupling
action has the special feature (which it shares with its electromagnetic
analogue~\cite{C97}) that its dependence on the metric $g_{\mu\nu}$ cancels out. Failure
to make proper allowance for the absence of this familiar simplifying property
in more general couplings, and in particular in couplings of gravitational and
dilatonic type, seems to be one of the main reasons why early evaluations
of the latter~\cite{DH89,CHH90} were so systematically erroneous.

The conditions on the other coupling term are more restrictive. In principle
the scalar source coefficient $\hat T$ might have various forms for diverse
scalar coupling theories that might be conceived, but in order for the
coupling to be describable as `` dilatonic'' in the sense associated most
particularly with the work of Dicke~\cite{D64}, it must be given by the trace
\be \hat T=\hat T_\mu{^\mu}\, ,\eqn{7}\fe
of the material stress energy tensor that is obtained from the primary
Lagrangian contribution ${\cal L}$ of the system according to the
usual geometric specification
\be\hat T{^{\mu\nu}}= 2\Vert g\Vert^{-1/2}
{\partial \big(\hat{\cal L}\Vert g\Vert^{1/2}\big)\over\partial g_{\mu\nu}}
= 2{\partial \hat{\cal L}\over\partial g_{\mu\nu}}
+\hat \Lambda g^{\mu\nu}\, .\eqn{8}\fe 

The reasonning whereby Dicke and other early workers (notably his
predecessor Jordan and his colleague Brans) were lead to a coupling of this
kind was based on the postulate that in a fully non linear treatment the
primary material action contribution would be given by an expression of the
form
$$ {\cal I}_{_{\rm D}}= \int{\cal L}_{_{\rm D}}\,
 \Vert g_{_{\rm D}}\Vert^{1/2}\, d^4 x \, ,$$
in which the Lagrangian density ${\cal L}_{_{\rm D}}$ depends on
whatever intrinsic material constituent fields may be necessary as well
as on a certain background metric $g_{_{\rm D}\mu\nu}$ say, that may
conveniently be referred to as the Dicke metric, in order to
distinguish it from the associated Einsten metric, $g_{_{\rm E}\mu\nu}$
say. The latter is characterised by the condition that the relevant
gravitational field action be proportional to the spacetime volume
integral of its Ricci tensor.  In a fully non linear treatment the
Dicke metric is related to the Einstein metric by a conformal
transformation of the form $g_{_{\rm D}\mu\nu}=e^{2\phi}g_{_{\rm
E}\mu\nu}$.  In a linearised treatment such as is used here, the
relevant Einstein metric can be taken to have the form $g_{_{\rm
E}\mu\nu}\simeq g_{\mu\nu}+h_{\mu\nu}$ where the ``unperturbed''
background metric $g_{\mu\nu}$ is strictly flat or at least Ricci flat,
and where $h_{\mu\nu}$ is the usual gravitational perturbation tensor.
The linearisation implies that one can also take $e^{2\phi}\simeq
1+2\phi$ and hence that the relevant Dicke metric can be taken to be
given in terms of the flat or Ricci flat background by a relation of
the form $g_{_{\rm D}\mu\nu}\simeq g_{\mu\nu}+h_{\mu\nu} +2\phi\,
g_{\mu\nu}$.

In the traditional kind of dilaton coupling theory theory (as envisaged
by Dicke~\cite{D64}), and also in the particular kind of low energy
classical limit of superstring theory that was considered by Dabholkar
and Harvey~\cite{DH89}, the $\phi$ dependence only comes in indirectly via the
dependence on $g_{_{\rm D}\mu\nu}$.  (Rather more complicated couplings occur
in some of the more eleborate models derived from superstring theory or
M-theory but, as remarked by Cho and Keum~\cite{CK98}, this does not
necessarily affect the form of their linearised weak field limits.)  Under
such conditions it follows that in the linearised limit we shall simply have
$${\cal L}_{_{\rm D}}\Vert g_{_{\rm D}}\Vert^{1/2}\simeq
 \Big({\cal L}+ T^{\mu\nu}(h_{\mu\nu}+2\phi\, g_{\mu\nu})\Big)
\Vert g\Vert^{1/2}$$
where ${\cal L}$ is obtained be substituting the unperturbed background metric
$g_{\mu\nu}$  in place of the Dicke metric $g_{_{\rm D}\mu\nu}$
in ${\cal L}_{_{\rm D}}$ and
$T^{\mu\nu}$ is the associated stress energy tensor as given by the 
usual formula (\ref{8}). The consequences of the purely gravitational
part of the coupling that arises in this  way have already been treated 
elsewhere~\cite{BC95,CB98,C98a,C98b}. The present work will be concerned
just with the dilatonic part, 
which will evidently be specified by a coupling term that reduces to 
the simple form $\hat T\phi$ as presented in (\ref{3}) with
$\hat T$ as given by (\ref{7}) and (\ref{8}).

Not much can be said about the equations of motion for the internal fields
characterising the material system until the form of its
primary  Lagrangian contribution $\hat
{\cal L}$ has been specified, but but it is evident quite generally that
independently of such details, in an empty background the equation of motion
for the axion field will be obtainable, using the gauge condition \be
\nabla^\mu B_{\mu\nu}=0\, \fe (the analogue of the Lorentz condition whose is
familiar in the context of electromagnetism) in the well known~\cite{VV87}
Dalembertian form
\be \nabla_{\!\sigma}\nabla^\sigma
B_{\mu\nu}=-4\pi m_{_{\rm J}}^{\,2}
 \hat W_{\mu\nu}\, .\eqn{10}\fe 
For the dilaton field no question of gauge arises at this stage: the relevant
field equation is thus immediately obtainable in the simple form
\be
\nabla_{\!\sigma}\nabla^\sigma \phi=-4\pi m_{_{\rm D}}^{\,-2} \,\hat T
\, .\eqn{11}\fe

\section{Distributional Sources}

As in the more familiar case of point particle models, or ``zero-branes'', the
problem of ultraviolet divergences for  higher dimensional ``$p$-branes'', and
in particular for string models, with $p$=1, because in these cases the
relevant source densities $\hat W{^{\mu\nu}}$, and $\hat T{^{\mu\nu}}$ are not
regular functions but Dirac type distributions that vanish outside the
relevant one or two dimensional world sheets, except of course in the case of
an ordinary medium, with the maximal space dimension, namely $p=3$, in an
ordinary 4-dimensional spacetime background of the kind to which the peresent
analysis is restricted.

In the case of general $p$-brane, with local $(p+1)$-dimensional worldsheet
imbedding given by $x^\mu=\overline x{^\mu}\{\sigma\}$ in terms of intrinsic
coordinates $\sigma^i$ ($i=0,1, ..., p$), so that the induced surface metric
will have the form \be \gamma_{ij}=g_{\mu\nu}\overline x{^\mu}{_{\! ,i}}\,
\overline x{^\nu}{_{\! ,j}} \, ,\eqn{12}\fe the relevant source distributions
will be expressible using the terminology of Dirac delta ``functions'' in the
form exemplified in the case of the vorticity flux by
\be \hat W{^{\mu\nu}}=\Vert g\Vert^{-1/2}\int \overline W{^{\mu\nu}}\, 
\delta^{4}[x-\overline x\{\sigma\}]\, \Vert\gamma \Vert^{1/2}
\, d^{p+1}\sigma\, ,\eqn{13}\fe
where $\Vert\gamma\Vert$ is the determinant of the induced metric (\ref{12}),
and where the where the surface vorticity flux bivector $\overline
W{^{\mu\nu}}$  is a {\it regular} antisymmetric tensor field on the worldsheet
(but undefined off it). The analogous expression for the stress momentum
energy source will be given by
\be \hat T{^{\mu\nu}}=\Vert g\Vert^{-1/2}\int \overline T{^{\mu\nu}}\, 
\delta^{4}[x-\overline x\{\sigma\}]\, \Vert\gamma \Vert^{1/2}
\, d^{ p+1}\sigma\, ,\eqn{14}\fe
where the  surface stress momentum energy density
$\overline T{^{\mu\nu}}$,  is a {\it regular} symmetric tensor field on the
worldsheet (but undefined off it).

The distributional nature of these source terms in cases for which
$p<3$ is a corollary of the similarly distributional nature of the
action density $\hat {\cal L}_{_{\rm ma}}$ as defined, 
according to (\ref{3a}),  to
include both the purely internal contribution $\hat{\cal L}$ and the
cross coupling contribution $\hat {\cal L}_{_{\rm co}}$. This
distributional action density will itself be expressible in the form
\be \hat {\cal L}_{_{\rm ma}}=\Vert g\Vert^{-1/2}\int \overline 
{\cal L}_{_{\rm ma}}\, 
\delta^{4}[x-\overline x\{\sigma\}]\, \Vert\gamma \Vert^{1/2}
\, d^{\rm p+1}\sigma\, ,\eqn{15}\fe
with
\be
\overline {\cal L}_{_{\rm ma}} = \overline{\cal L}+\overline 
{\cal L}_{_{\rm co}}\eqn{16}\fe
where the primary contribution or ``master function'' 
$\overline{\cal L}$ and the secondary coupling contribution
 $\overline {\cal L}_{_{\rm co}}$ are both well behaved scalar
functions on the string worldsheet but undefined off it. Of these, the master
function $\overline{\cal L}$  will be the intrinsic worldsheet Lagrangian,
defined as a function just of the relevant internal, worldsheet confined,
fields, such as currents, on the string, and of its induced metric, while the
cross coupling contribution will be given in terms of the worldsheet confined
fields $\overline W{^{\mu\nu}}$ and $\overline T{^{\mu\nu}}$ by
\be 
\overline {\cal L}_{_{\rm co}}= 
{_1\over^2}\overline W{^{\mu\nu}}B_{\mu\nu}
+\overline T \phi\, ,\eqn{17} \fe
where $\overline T=\overline T_\mu{^\mu}$. 

In terms of these well behaved worldsheet functions, the
corresponding material contribution (\ref{2}) to the action
can be expressed directly, without recourse to heavy distributional
machinery, as a simple (p+1)-surface integral in the form
\be {\cal I}_{_{\rm ma}}=\int \overline {\cal L}_{_{\rm ma}}
\, \Vert \gamma\Vert^{1/2}\, d^{\rm p+1}\sigma\, .
\eqn{18}\fe
In particular, the regular surface stress energy tensor $\overline
T{^{\mu\nu}}$  needed for the purpose of applying (\ref{17}) is obtainable
directly from the world sheet master function, without the use of
distributions, using the formula 
\be
\overline T{^{\mu\nu}}= 2\Vert\gamma\Vert^{-1/2} {\partial
(\overline{\cal L}\Vert\gamma\Vert^{1/2})\over\partial g_{\mu\nu}}\,
.\eqn{19}\fe

As was remarked above, the errors in the early literature on this
subject\cite{DH89,CHH90} were largely attributable to failure to take proper
account of the fact that unlike what occurs in the historically familiar
special cases of electromagnetic and axionic coupling, for more general cases
such as those of gravitational and dilatonic coupling the relevant coupling
action contribution will be metric dependent. In order to allow for 
this, it is useful to work out the appropriately constructed hyper-Cauchy
tensor (a relativistic generalisation of the Cauchy elasticity tensor of
classical mechanics) which is defined by
\be 
\overline{\cal C}{^{\mu\nu\rho\sigma}}=\Vert\gamma\Vert^{-1/2}
{\partial\over\partial g_{\mu\nu}}\Big(\overline T{^{\rho\sigma}}
\Vert\gamma\Vert^{1/2}\Big) \, ,
\eqn{20}\fe
or equivalently, in manifestly symmetric form by
\be\overline{\cal C}{^{\mu\nu\rho\sigma}}=\overline{\cal C}{^{\rho\sigma\mu\nu}}
=2\Vert\gamma\Vert^{-1/2}{\partial^2
\big(\overline{\cal L}\Vert\gamma\Vert^{1/2})\over\partial g_{\mu\nu}
\partial g_{\rho\sigma}}\, . \eqn{21}\fe

\section{Brane worldsheet geometry}

If the relevant radiation fields $B_{\mu\nu}$, and $\phi$ are considered to be
regular background fields attributable to external souces, the treatment of a
brane system of the kind described in the preceeding section will be
straightforeward, but it is evident that this will not be the case for the
radiation fields produced by the brane itself, since they will be singular
just where their evaluation is needed. 

Even for the treatment just of the regular case in which the relevant
radiation fields are of purely external origin, and {\it a fortiori} for the
treatment of the more delicate problem of self interaction, it is desirable
-- before proceeding to the derivation of the dynamical equations that ensue
from the action (\ref{18}) to recapitulate the essential geometric
concepts\cite{C92,C97a} that are needed for the kinematic description 
of the evolving worldsheet. (The unavailability of this machinery at the time of the
pioneering work\cite{DH89,CHH90} on the Goto-Nambu case is one of the reasons
for the use of the misleading shortcut methods responsible for the confusion
that beset this subjet before the recent clarification~\cite{CB98,BD98a} ).

Point particle kinematics can conveniently be developed in terms of the
timelike worldline tangent vector $u^\mu$ that is uniquely fixed by the
condition of being future directed with unit normalisation, $u^\mu 
u_\nu=-1$, and of the associated acceleration vector that is given
in terms of covariant differentiation with respect to the (flat or curved)
spacetime background metric $g_{\mu\nu}$ by $a^\mu=u^\nu\nabla_{\!\nu} u^\mu$.
For higher brane cases, and in particular for the strings with which we shall be
concerned here, a less specialised kinematic description must be used.
Instead of the unique tangent vector $u^\mu$ and the derived vector
$a_\mu$ that suffice for the ``zero brane'' case, the kinematic
behaviour of higher ``branes'' starting with the case of strings (i.e.
``one branes'') is most conveniently describable~\cite{C92,C97a} in
terms of its first and second fundamental tensors. The former is 
definable as tangential projection tensor $\eta^\mu_{\,\nu}$ say,
which obtained by index lowering from the spacetime background
projection of the inverse of the induced metric as given by the formula
\be \eta^{\mu\nu}=\gamma^{ij}\overline x{^\mu}{_{\! ,i}}\, 
\overline x{^\mu}{_{\! ,j}}\, ,\eqn{22}\fe 
This {\it first fundamental} tensor can conveniently be used to rewrite
the expressions (\ref{19}) and (\ref{20}) in the more practical forms 
\be
\overline T{^{\mu\nu}}= 
2{\partial\overline{\cal L}\over\partial g_{\mu\nu}}
+\overline\Lambda \eta^{\mu\nu}\, .\eqn{23}\fe
and
\be 
\overline{\cal C}{^{\mu\nu\rho\sigma}}
 ={\partial \overline T{^{\rho\sigma}} \over \partial
g_{\mu\nu}}+{_1\over^2}\overline T{^{\rho\sigma}} \eta^{\mu\nu}\, ,
\eqn{24}\fe

The corresponding {\it second fundamental tensor} $K_{\mu\nu}{^\rho}=
K_{\nu\mu}{^\rho}$, is obtained from the first fundamental tensor using
the tangentially projected differentiation operator
\be \overline\nabla_{\!\mu} =
\eta_\mu^{\ \nu}\nabla_{\!\nu}\, ,\eqn{25} \fe
according to the prescription
\be K_{\mu\nu}{^\rho}=\eta^\sigma_{\ \nu}\overline\nabla_{\!\mu}\eta^\rho_{\ 
\sigma}\, .\eqn{26}\fe
The condition of integrability of the worldsheet is the Weingarten identity,
to the effect that this second fundamental tensor should be symmetric
on its first two indices, i.e. 
\be K_{[\mu\nu]}{^\rho}=0\, .\eqn{27}\fe
This tensor has the noteworthy
property of being worldsheet orthogonal on its last index, but tangential
on its (by the Weingarten identity interchangable) first pair of indices,
\be K_{\mu\nu}{^\sigma}\eta_\sigma{^\rho}=0
=\perp^{\!\lambda}_{\,\mu}K_{\lambda\nu}{^\rho}=0\, ,\eqn{28}\fe
 using the notation
\be \perp^{\!\mu}_{\,\nu}=g^\mu{_\nu}-\eta^\mu{_\nu} \eqn{29}\fe
for the tensor of projection orthogonal to the worldsheet.

Whereas the full second fundamental tensor is needed for dealing with
gravitational coupling~\cite{BC95,CB98}, the treatment of the simpler cases
considered here requires only its trace, namely the curvature vector
\be K^\rho=K_\mu{^{\mu\rho}}=\overline\nabla_{\!\nu}\eta^{\nu\rho}\, ,
\eqn{30}\fe 
which must evidently be worldsheet orthogonal, i.e. 
\be \eta^\rho{_\sigma}K^\sigma=0\, .\eqn{31}\fe

In terms of the background Riemann Christoffel connection 
$\Gamma_{\mu\ \rho}^{\,\ \nu}=g^{\nu\sigma}\big(g_{\sigma(\mu,\rho)}
-{_1\over^2}g_{\mu\rho,\sigma}\big)$ this curvature vector will be
expressible in explicit detail as
\be K^\nu={1\over\sqrt{\Vert \gamma\Vert}}\Big(
\sqrt{\Vert\gamma\Vert}\gamma^{ij}\overline x^\nu_{\, ,i}\Big){_{,j}}+
\gamma^{ij}\overline x^\mu_{\, ,i}\overline x^\rho_{\, ,j}\Gamma_{\mu\ \rho}^{\,\ \nu}\, .
\eqn{32}\fe

In the  particulary simple case a Dirac membrane or a Nambu Goto string (i.e.
one for which the master function $\overline{\cal L}$ is just a constant) that
is free, in the sense that it is not subjected to any external force, the ``on
shell'' configurations (i.e. the solutions of the variational dynamical
equations) will simply be characterised by the condition that the vector
(\ref{32}) vanishes, $K^\mu=0$, but this simple vanishing curvature condition
will not be satisfied for more general models (such as those needed for
superconducting strings~\cite{C89,CP94,GPB98}) nor when dilatonic and 
axionic forces are involved as in the cases analysed in the present work. 

It is to be remarked that the orthogonality property (\ref{31}) of the 
curvature vector is to be contrasted with the tangentiality property of the
stress energy tensor,
\be \perp^{\!\lambda}_{\,\mu}\overline T{^{\mu\nu}}=0\, ,\eqn{33} \fe
and of the hyper-Cauchy tensor,
\be \perp^{\!\lambda}_{\,\mu}\overline {\cal C}{^{\mu\nu\rho\sigma}}
=0\, .\eqn{34}\fe 

Subject to the requirement that the worldsheet supported field
$\overline W{^{\mu\nu}}$ be constructed from internal worldsheet fields in
such a way as to aquire the corresponding tangentiality property
\be \perp^{\!\lambda}_{\,\mu}\overline W{^{\mu\nu}}=0\, ,\eqn{35} \fe
it can be seen that the corresponding distributional conservation
law (\ref{6}) can be equivalently expressed
in terms of tangentially projected differentiation
as the regular worldsheet flux conservation law
\be \overline\nabla_{\!\mu}\overline W{^{\mu\nu}}=0\, .\eqn{36}\fe

\section{The force density formula}

For the purpose of the deriving the equations of motion of the material system
from the variation principle, the most general variations to be considered are
perturbations of the relevant internal fields, which have not yet been
specified, and infinitesimal displacements with respect to the background
characterised by the metric $g_{\mu\nu}$ and the the linearly coupled
axionic and dilatonic fields.

The effect of displacements can be conveniently analysed using
a Lagrangian treatment in which not just the internal coordinates
$\sigma^i$ but also the background coordinates $x^\mu$ are considered
to be dragged along by the displacement, so that the relevant field 
variations are given just by the corresponding Lie derivatives with respect 
to the displacement vector field $\xi^\mu$ under consideration.
This leads to the formulae
\be \delta B_{\mu\nu}=\xi^\sigma\nabla_{\!\sigma} B_{\mu\nu}-2
B_{\sigma[\mu}\nabla_{\!\nu]}\xi^\sigma \eqn{37} \fe
for the axionic field, and 
\be \delta \phi=\xi^\sigma\nabla_{\!\sigma}\phi \eqn{38} \fe
for the dilatonic field,
while finally for the background metric itself one has the well known
formula
\be \delta g_{\mu\nu}=2\nabla_{\!(\mu}\xi_{\nu)}\, .\eqn{39} \fe

The postulate that the internal field equations are satisfied means that
perturbations of the relevant internal fields have no effect on the action
integral ${\cal I}_{_{\rm ma}}$, with the  implication that for the purpose of
evaluating the variation $\delta{\cal I}_{_{\rm ma}}$ there will be no loss of
generality taking the variations of these so far unspecified internal fields
simply to be zero. This means that the only contribution from the
first term in (\ref{16}) will the one provided by the background metric
variation, for which we obtain the familiar formula
\be  \delta\Big(\Vert\gamma\Vert^{1/2}\overline{\cal L}\Big)=
{_1\over^2}\Vert\gamma\Vert^{1/2}\,
\overline T{^{\mu\nu}}\delta g_{\mu\nu}\, .\eqn{40}\fe

The worldsheet flux conservation law (\ref{36}) is
interpretable~\cite{C92} as meaning that  $\overline W{^{\mu\nu}}$ is related
by Hodge type duality to the exterior derivatives of corresponding  worldsheet
differential forms (that will generically be of order p-2 respectively). In
all the usual applications (including the continuuous medium example
mentionned in the preceeding section and the string case developed below)
these differential forms will be included among (or depend only on) the
relevant independently variable internal fields whose variation can be
taken to be zero for the purpose of evaluationg $\delta{\cal I}_{_{\rm ma}}$
when the internal field equations are satisfied. This means
that the variation of the bivector surface density will also vanish, i.e. we
shall have
\be \delta\Big(\Vert\gamma\Vert^{1/2} \overline W{^{\mu\nu}}\Big)=0\, 
. \eqn{41} \fe
It follows that the axionic contribution from (\ref{17}) to the variation of
the integrand in (\ref{18}) will be given simply by 
\be \delta\Big(\Vert\gamma\Vert^{1/2}\big( 
{_1\over^2}\overline W{^{\mu\nu}} B_{\mu\nu}\big)\Big)=
\Vert\gamma\Vert^{1/2}\big( {_1\over^2}\overline W{^{\mu\nu}}
\delta B_{\mu\nu}\big)
\, .\eqn{42}\fe
 
The systematic absence of any contribution from the background metric
variation $\delta g_{\mu\nu}$ to action variation terms such as this, not only
in the axionic case considered here but also in its more widely familiar
analogue, sets a potentially misleading precedent that encourages a dangerous
tendency (one of the main sources of error in earlier work~\cite{DH89,CHH90})
to forget to check the possibility of metric variations in more general
contexts. Although it does not contribute to the axionic term (\ref{42})
allowance for the background metric variation (\ref{39}) turns out to be of
paramount importance not only in the gravitational case~\cite{BC95,CB98} but
also for the evaluation of the dilatonic contribution with which we are
concerned here. It can be seen from (\ref{24}) that we shall have
 \be \delta\Big(\Vert\gamma\Vert^{1/2}\
\overline T \Big)= \Vert\gamma\Vert^{1/2}\big( \overline T{^{\mu\nu}}+
 \overline{\cal C}{^{\mu\nu}}\big) \,\delta g_{\mu\nu}
 \, ,\eqn{43}\fe
using the notation
\be   \overline{\cal C}{^{\mu\nu}} = \overline{\cal C}{^{\mu\nu\rho}}_\rho
\, ,\eqn{44}\fe
for the trace of the hyper Cauchy tensor. Thus  despite its deceptively simple
scalar nature, the dilatonic coupling gives rise to a corresponding
contribution that works out to be given by an expression of the not quite
trivially obvious form
 \be \delta\Big(\Vert\gamma\Vert^{1/2}\
\overline T \phi\Big)= \Vert\gamma\Vert^{1/2}\Big(
 \overline T \delta\phi + (\overline T{^{\mu\nu}}+
 \overline{\cal C}{^{\mu\nu}} )\phi \,\delta g_{\mu\nu}\Big)
 \, .\eqn{45}\fe

To evaluate the integrated effect of the contributions (\ref{40}), (\ref{42}),
and (\ref{45}) the next step is the routine procedure of substitution of the
relevant Lie derivative formulae (\ref{37}), (\ref{38}),  and (\ref{39})
followed by absorbtion of the terms involving derivatives of the displacement
fields into pure worldsheet current divergences. For the primary contribution
given by (\ref{40}) one thereby obtains an expression of the familiar form
\be {_1\over^2}\overline T{^{\mu\nu}}\delta g_{\mu\nu} =-\xi^\mu
\overline\nabla_{\!\nu}\overline T{^\nu}_\mu+\overline\nabla_{\!\mu}
\big(\xi^\nu \overline T{^\mu}_\nu\big)\, .\eqn{46}\fe
while the corresponding expression for
axionic coupling contribution (\ref{42}) will simply be given by
\be {_1\over^2}\overline W{^{\mu\nu}}\delta B_{\mu\nu}=
{_1\over ^2} \xi^\mu N_{\mu\nu\rho} \overline W{^{\nu\rho}} 
+\overline \nabla_{\!\mu}\big(\xi^\nu B_{\nu\rho}\overline W{^{\mu\rho}}
\big)\, .\eqn{47}\fe
However the dilatonic coupling contribution (\ref{45}) is not so simple: in
addition to the obvious scalar field variation contribution given by
\be\overline T\delta\phi=\xi^\mu\overline T\nabla_{\!\mu}\phi\, .\eqn{48}\fe
there will be another less obvious contribution (the one that tended to be
overlooked in earlier work) given by
\be  \big(\overline T{^{\mu\nu}}+\overline{\cal C}{^{\mu\nu}}\big)\phi\,
\delta g_{\mu\nu}=-\xi^\mu\overline\nabla_{\!\nu}\Big(2
\big(\overline T{^\nu}_\mu+ \overline{\cal C}{^\nu}_\mu\big)\phi\Big)+
\overline\nabla_{\!\mu}\Big(2\xi^\nu\big(\overline T{^\mu}_\nu+
\overline{\cal C}{^\mu}_\nu\big)\phi \Big)\, .\eqn{49}\fe

When these expressions are used to evaluate the variation of the
action integral (\ref{15}) due to a displacement confined to a bounded
region of the worldsheet, the application of the relevant (p+1) dimensional
version of Green's theorem removes the contributions from the divergence
terms, so that one is left with an expression of the standard form
\be \delta{\cal I}_{_{\rm ma}}=\int \xi^\mu\big( \overline f_\mu
-\overline\nabla_{\!\nu}\overline T{^\nu}_\mu 
\big) \Vert \gamma\Vert^{1/2}\, d^{p+1}\sigma\, ,\eqn{50} \fe
so that the application of the variation principle gives
the corresponding dynamical equation
\be \overline\nabla_{\!\nu}\overline T{^{\mu\nu}}= \overline f{^\mu}
\eqn{51} \fe
in which the vector $\overline f{^\mu}$ represents the total
force density exerted by the various radiation fields involved.
Using the foregoing expressions, this force density can immediately be read
out in the form
\be \overline f{^\mu}=\overline f_{_{\!\rm J}}{^\mu}+\overline f_{_{\!\rm D}}
{^\mu}\, ,\eqn{52}\fe
in which the axionic contribution can be seen from (\ref{47}) to be given by
the well known formula (the axionic analogue of the Lorentz force formula
in electromagnetism) given by
\be\overline f_{_{\!\rm J}}{^\mu}={_1\over^2}N^\mu{_{\nu\rho}}
\overline W{^{\nu\rho}} \eqn{53}\, ,\fe
which seemed unfamiliar~\cite{VV87} when first derived in the present context,
but which is in fact interpretable just as the immediate relativistic
generalisation of the historic formula on which the theory of flying is based,
namely the Joukowski force law for the lift (due to the Magnus effect) on a
long thin (i.e. string-like) aeroplane wing. What is not so well known
is the corresponding formula for the dilatonic force density,
which can be seen from (\ref{48}) and (\ref{49}) to be given by
\be \overline f_{_{\!\rm D}}{^\mu} =\overline T\nabla^\mu\phi
-\overline\nabla_{\!\nu}\Big(2
\big(\overline T{^{\mu\nu}}+
\overline{\cal C}{^{\mu\nu}}\big)\phi\Big) \, .\eqn{54}\fe

\subsection{The force on a Nambu-Goto string.} 

The proceeding formulae apply to domain wall type membrane models (with $p=2$)
as well as to simple point particle models (with $p=0$), but from this stage
onwards we shall restrict our attention to the case of string models, as
characterised by $p=1$. Before further restricting attention to the very
special case of Nambu-Goto type string models, it is worthwhile to
recapitulate some relevant features that are shared by more general string
models, including the kind needed~\cite{C89,CP94,GPB98} for describing the
effects of the type of supercontivity whose likely occurrence in cosmic
strings was originally predicted by Witten~\cite{W85}. 

In the higher dimensional branes the vorticity flux $\overline W{^{\mu\nu}}$
might depend on internal field variables of the model, while for a point
particle model no such source can exist at all. In the intermediate case of a
string there is no obstruction to the existence of a vorticity flux but it
cannot depend on any internal field variables of the model: the only way the
conservation law (\ref{36}) can be satisfied on a 2-dimensional worldsheet is
for the vorticity flux to have the form \be \overline
W^{\mu\nu}=\bar\kappa{\cal E}^{\mu\nu}\, ,\eqn{55}\fe where ${\cal
E}^{\mu\nu}$ is the antisymmetric unit surface element tensor and $\bar\kappa$
is a constant that is interpretable as representing the momentum circulation
around the string of the Zel'dovich type  fluid representing the axion field
-- which means that it will be an integral multiple of Planck's constant, i.e.
an integral multiple of $2\pi$ in the unit system we are using with $\hbar$
set to unity. Using the traditional dot and dash notation $\dot
x^\mu=\overline x{^\mu_{\, ,_0}}$ and $ x{^{\prime\mu}}= \overline x{^\mu_{\,
,_1}}$ for the effect of partial differentiation with respect to worldsheet
coordinates $\sigma^{_0}$ and $\sigma^{_1}$ the antisymmetric unit surface
element tensor will be given by
\be {\cal E}^{\mu\nu}=
2\big(\Vert\gamma\Vert\big)^{-1/2}\,x^{[\mu}_{\ \, ,_0}x^{\nu]}_{\ ,_1}
\, ,\eqn{56}\fe

In the case of a string, the fundamental tensor will be given in terms
of this unit tangent bivector by
\be \eta^\mu_{\ \nu}={\cal E}^\mu_{\ \rho}{\cal E}^\rho_{\ \nu}\, .\eqn{57}\fe
One of the reasons why the kind tensorial analysis used here was not developed
much earlier for the purpose of application to string dynamics is that the
heavy algebra involved in the use of coordinate dependent expressions such as
that on the right hand side of the curvature formula (\ref{32}) could be
alleviated to some extent by the use of specialised internal coordinate
systems of the conformal type characterised by the conditions 
\be \dot x^\mu
x^\prime_{\, \mu}=0\, \hskip 1 cm 
\dot x^\mu\dot x_\mu+x^{\prime\mu}x^\prime_{\,\mu}=0\, ,\eqn{58}\fe
which imply that the relation 
\be {\Vert \gamma\Vert}^{1/2}=
x^{\prime\mu}x^\prime_{\,\mu}=-\dot x^\mu\dot x_\mu\, ,\eqn{59}\fe
If  $\sigma^{_0}$ and $\sigma^{_1}$ are restricted to
satisfy these conditions (which are frequently incompatible with other
compelling desiderata) then the unweildy formula (\ref{32}) for the curvature
vector that plays such an important role will be replaceable by the handier
expression
\be K^\nu=\Vert\gamma\Vert^{-1/2}\Big(
x^{\prime\prime\nu}-\ddot x^\nu + (x^{\prime\mu}x^{\prime\rho} -\dot
x^\mu\dot x^\rho)\Gamma_{\mu\ \rho}^{\,\ \nu}\Big)\, .\eqn{60}\fe
If the background metric $g_{\mu\nu}$ is not just empty but actually flat then
this formula will be further simplifiable by elimination of the final term if
one is willing to restrict the background cordinates to be of Minkowski type,
but of couse such a restriction may not be what is most convenient for
exploiting symmetries, such as occur in circular string loop configurations
for which spherical or cylindrical coordinates might be preferable. The
development of string dynamics has been unnecessarily delayed by over reliance
on the special gauge characterised by Minkowski coordinates on the background
and conformal coordinates on the worldsheet, rather that using the kind of
geometrical approach followed here, which provides more elegant and concise
formulae for general purposes. This more powerful geometric approach is of
course particularly advantageous for applications in which for various
technical reasons the usual (conformal cum Minkowski) kind of gauge may be
unsuitable.

From this point on, attention will be restricted to the special simple case of
string models of Nambu-Goto type, which includes the case that arises in the
low energy limit of string theory considered by Dabholkar and
Harvey~\cite{DH89}. Such models are characterised by the condition that the
relevant master function $\overline{\cal L}$ is simply a constant, which means
that it will be expressible in the form
\be \overline{\cal L}= -m_{_{\rm K}}^{\,2} \, ,\eqn{601}\fe
where $m_{_{\rm K}}$ is a fixed mass scale that will be referred to as
the Kibble mass to distinguish it from other mass scales in the
theory.  In the context of cosmic string theory it
is generally expected that it should be of the same order of magnitude
as the Higgs mass, $m_{_{\rm X}}$ say, that is associated with the
underlying ``spontaneously broken'' symmetry of the vacuum.
In the context of superstring theory the quantity $m_{_{\rm
K}}$ is usually supposed to be of the order of magnitude of the Planck
mass $m_{_{\rm P}}$. 

The other parameters needed to complete the specification of the theory are
the quantities $m_{_{\rm J}}$, $m_{_{\rm D}}$ and $\overline\kappa$ that have
already been introduced. To match the present formulation to the equivalent
low energy linearised limit theory in the slightly different notation used by
Buonanno and Damour~\cite{BD98a} their  parameters $\alpha$, $\lambda$, $\mu$
are identifiable as being given by the relations $\alpha=m_{_{\rm P}}/m_{_{\rm
D}}$, $\lambda=\overline\kappa m_{_{\rm P}} m_{_{\rm J}}/2$ and $\mu=m_{_{\rm
K}}^{\,2}$. The special values corresponding to the low energy superstring
theory limit considered by  Dabholkar  and Harvey are given by $\ \alpha=1\ $,
$\ \lambda=\mu\ $, which in the formulation used here is equivalent to the
conditions
\be m_{_{\rm D}}=m_{_{\rm P}}\, \hskip 1 cm 2m_{_{\rm K}}^2=\overline\kappa
m_{_{\rm J}}m_{_{\rm P}}\, .\eqn{61}\fe

Whether or not the particular conditions (\ref{61}) are satisfied (and of
course quite independently of whether the internal coordinate gauge satisfies
the conditions  (\ref{58}) on which the specialised formulae (\ref{59}) and
(\ref{60}) depend) the surface stress momentum energy tensor of the string can
be seen from (\ref{23}) to be proportional to the fundamental tensor,
according to the formula
\be \overline T{^{\mu\nu}}=-m_{_{\rm K}}^{\,2}\eta^{\mu\nu} \, ,\eqn{62}\fe
and so its trace will be given by
\be \overline T=-2 m_{_{\rm K}}^{\,2}\, .\eqn{63}\fe
The corresponding the hyper-Cauchy tensor is obtainable\cite{BC95} from
(\ref{24}) in the form
\be \overline{\cal C}{^{\mu\nu\rho\sigma}}=m_{_{\rm K}}^{\, 2}
\big(\eta^{\mu(\rho}\eta^{\sigma)\nu}-\textstyle{1\over 2}
\eta^{\mu\nu}\eta^{\rho\sigma}\big)\, .\eqn{64}\fe
It can be seen from this that in this special Nambu-Goto case the trace tensor
that appears in the dilatonic force formula (\ref{54}) will vanish, i.e. one
obtains
\be \overline{\cal C}{^{\mu\nu}}=0\, .\eqn{65}\fe
It is to be emphasised that that this simplification is a special feature of
the string case, and that it does not hold in the higher dimensional case of a
Dirac membrane, nor even in the trivial lower dimensional case of a point
particle with mass $m$ and unit 4-velocity vector $u^\mu$, for which one
obtains $\overline{\cal C}{^{\mu\nu}}=-m u^\mu u^\nu/2$.

It follows from (\ref{53}) and (\ref{55}) that the Joukowsky force density
exerted by the axionic fluid on a string (of any kind) will be given  by
\be\overline f_{_{\!\rm J}}{^\mu}={_1\over^2}\overline\kappa
N^\mu{_{\nu\rho}}{\cal E}{^{\nu\rho}}\, . \eqn{66}\fe
It follows from (\ref{54}) using (\ref{62}) and (\ref{65}) that in the case of
a Nambu Goto string the corresponding dilatonic force density contribution
will be obtainable -- with the aid of the defining formula (\ref{30}) for the
curvature vector $K^\mu$ -- in the form
\be\overline f_{_{\!\rm D}}{^\mu} =2m_{_{\rm K}}^{\,2}\big(
\phi K^\mu- \perp^{\mu\nu}\!\nabla_{\!\nu}\phi\big)\, . \eqn{67}\fe
Simple though it is, this formula does not seem to have been previously
made available in the literature.

It is to be observed that -- as needed to avoid overdetermination in the
Goto-Nambu case -- the force contributions (\ref{66}) and (\ref{67}) are both
identically orthogonal to the string worldsheet. It is evident that if the
dilatonic field were due only to high frequency radiation from an external
source then the first term on the right in (\ref{67}) would be relatively
negligible, but it will be shown below that this term is not at all negligible
for a dilatonic field due to self interaction.

\section{ Allowance for regularised self interaction}

In cases where self interaction is involved, it is commonly
convenient to decompose the relevant linear interaction fields --
which in the present instance are $B_{\mu\nu}$ and $\phi$ -- into
a short range contribution determined via the relevant Green function
by the immediately neighbouring source distribution, and a longer range
residual contribution that includes allowance for incoming radiation
from external sources. More particularly, in the present instance,
it will be useful to consider the relevant fields 
$B_{\mu\nu}$ and $\phi$ the sums of short range contributions that will be
indicated by a widehat and long range parts that will be indicated by
a widetilde in the form
\be
B_{\mu\nu}=\widetilde B_{\mu\nu}+\widehat B_{\mu\nu}\, ,
\hskip 1 cm \phi=\widetilde \phi+\widehat \phi\, ,\eqn{68}\fe
This will evidently give rise to corresponding decompositions 
\be f_{_{\!\rm J}}^{\,\mu}=\widetilde f_{_{\!\rm J}}^{\,\mu}
+\widehat f_{_{\!\rm J}}^{\,\mu} \, ,\hskip 1 cm 
f_{_{\!\rm D}}^{\,\mu}=\widetilde f_{_{\!\rm D}}^{\,\mu}
+\widehat f_{_{\!\rm D}}^{\,\mu}
 \eqn{69}\fe
for the asociated force densities as specified by the general formulae 
(\ref{53}), (\ref{54}) or their Nambu Goto string specialisations
 (\ref{66}), (\ref{67}).

In many contexts the coupling is so weak that the local self force
contributions $\widehat f_{_{\!\rm J}}^{\,\mu}$ and $\widehat f_{_{\!\rm D}}
^{\,\mu}$ can be neglected. However in cases for which one needs to take
account of the self induced contributions $\widehat B_{\mu\nu}$ and $\widehat
\phi$, there will be difficulties arising from the fact that the relevant
source fields on the right hand sides of the field equations (\ref{11}) and
(\ref{11}) will not be the regular worldsheet supported tensor fields
$\overline W{^{\mu\nu}}$, and $\overline T$, but the corresponding
4-dimensionally supported distributions, $\hat W{^{\mu\nu}}$, and
$\hat T$ as constructed according to the prescriptions (\ref{13})  and
(\ref{14}). For sources such as these, the resulting field contributions will
diverge in the thin worldsheet limit.

As in the familiar point particle case, so also for a string, one can obtain
an appropriately regularised result by supposing that (as will be entirely
realistic in cases such as that of a cosmic string model for a vortex defect of
the vacuum) the underlying that the physical system one wishes to describe is
not quite infinitely thin but actually has finite spacial extent that can be
used to specify an appropriately microscopic ``ultraviolet'' cut off length
scale $\delta_\ast$ say. This will be sufficient for regularisation in the
point particle case, but in the string case it will also be necessary to
introduce a long range ``infrared'' cut off length scale $\Delta$ say that
might represent the macroscopic mean distance between neigbouring strings.  In
the case of a string an an ordinary 4-dimensional spacetime background, it can
be seen (following the example~\cite{C97} of the electromagnetic prototype
considered by Witten in his original discussion~\cite{W85} of
``superconducting'' cosmic strings) that the dominant contribution to relevant
Green function integrals in the ultraviolet limit  will then be proportional
to a logarithmic regularisation factor of the form
\be \widehat l= {\rm ln}\big\{ {\Delta^2 /\delta_\ast^{\, 2} } \big\}\, .
\eqn{70}\fe
More specifically, the dominant contribution to the axionic self field
arising from the Dalembertian source equation (\ref{10}) will be given by
\be\widehat B_{\mu\nu}=\widehat l\ m_{_{\rm J}}^{\, 2}
\overline W_{\mu\nu}\, ,\eqn{71}\fe
with $\overline W_{\mu\nu}$ given by (\ref{55}),
while similarly the dominant contribution to the dilatonic self field
arising from the Dalembertian source equation (\ref{11}) will be given by
\be\widehat\phi=\widehat l\, m_{_{\rm D}}^{\,-2}\, \overline T\, .\eqn{73}\fe
(If the microscopic  axial current source distribution were very different
from that of the stress energy trace source for the dilatonic field, the
natural cut off $\delta_\ast$ that would be most appropriate for the latter
might differ somewhat from what would apply to the former, but the effect of
such a difference could be considered as a higher order correction that need
not be taken into account so long as we are only concerned with the dominant
contribution.)

For the purposes of substitution in the force formulae -- (\ref{53}),
(\ref{54}) or their Nambu Goto string specialisations (\ref{66}), (\ref{67})
-- knowledge just of the regularised self fields  $\widehat B_{\mu\nu}$ and
$\widehat\phi$ is not immediately sufficient. The problem is that these
regularised values are well defined only on the worldsheet, which means that
they do not directly provide what is needed for a direct evaluation of the
gradients that are required: there is no difficulty for the terms involving
just the tangentially projected gradient operator $\overline\nabla_{\!\nu}$
but it can be seen that there are also contributions from the unprojected
gradient operator $\nabla_{\!\nu}$ which is directly meaningfull only when
acting fields whose support extends off the worldsheet.

It fortunately turns out that this problem has a very simple general solution,
of which particular applications in particular gauges are implicit in  
much previous work \cite{DQ90a,DQ90b,QS90,CHH90,K93,BS95,BS96}  
and which I formulated explicitly in conveniently covariant and more generally
utilisable form in the specific context of the electromagnetic
case~\cite{C97}. What one finds -- by examining the string worldsheet limit
behaviour of derivatives of the relevant Dalembertian Green function -- is
that the appropriate regularisation of the gradients on the string worldsheet
is obtainable simply by replacing the ill defined operator $\nabla_{\!\nu}$ by
by the corresponding regularised gradient operator
\be\widehat\nabla_{\!\nu}=\overline\nabla_{\!\nu}+{_1\over ^2}K_\nu \,
,\eqn{74}\fe 
where $K^\mu$ is the worldsheet curvature vector that is defined by the
formula (\ref{30}) and that is expressible, if one is willing to allow oneself
to be restricted to the use of conformal worldsheet coordinates, by a more
detailed prescription of the form (\ref{60}). In the explicit application of
the formula (\ref{74}), it is sufficient, in the case of a scalar field
$\varphi$, to use the simple expression
$\overline\nabla{^\mu}\varphi=\gamma^{ij}\overline x^\mu{_{,i}}\varphi_{,j}$
for the tangentially projected gradient, but for a tensorial field  there will
also be contributions depending on the background Riemann Christoffel
connection $\Gamma_{\mu\ \rho}^{\,\ \nu}$, which is also involved in the
detailed expressions (\ref{32}) and (\ref{60}), unless one is using Minkowski
coordinates in a flat spacetime backckround.

Applying this procedure to the axionic Kalb Ramond 2-form, one sees from
(\ref{5}) and (\ref{71}) that the corresponding regularised local current
3-form contribution will be given by 
\be\widehat J_{\mu\nu\rho}=3\widehat\nabla_{\![\mu}\widehat B_{\nu\rho]}
={_1\over^2}\widehat l\,\big(6\overline\nabla_{\![\mu}\overline W_{\nu\rho]}
+3K_{[\mu}\overline W_{\nu\rho]}\big) \, .\eqn{75}\fe
Taking account of the surface flux conservation law (\ref{36}), and
using the defining relation (\ref{30}) for the curvature vector $K^\mu$,
it can be seen that the axionic force density (\ref{53}) will be expressible
as a world sheet divergence in the form
\be\widehat f_{_{\!\rm J}}^{\,\mu}=-\overline\nabla_{\!\nu}
\widehat T_{_{\!\rm J}}{^{\mu\nu}}
\, ,\eqn{76}\fe
in terms of a regularised local axionic stress energy tensor given by
\be \widehat T_{_{\!\rm J}}{^{\mu\nu}}=\widehat B{^\mu}_\rho \overline
W{^{\nu\rho}} -{_1\over^4}\widehat B_{\rho\sigma}
\overline W{^{\rho\sigma}}\eta^{\mu\nu} \, ,\eqn{77}\fe
which can be seen from  (\ref{55}) and (\ref{71}) to reduce with the aid of
(\ref{57}) to the simple explicit form
\be\widehat T_{_{\!\rm J}}{^{\mu\nu}}=-{_1\over ^2}\widehat l\
\bar\kappa^2 m_{_{\rm J}}^{\, 2}\eta^{\mu\nu}\, ,\eqn{78}\fe
which is evidently isotropic with respect to the 2-dimensional worldsheet
geometry, like the intrinsic stress energy tensor in the  Nambu-Goto case

When one applies same procedure to the dilatonic self force contribution in
(\ref{54}) one finds that it too can be formulated as a world sheet divergence
in the analogous form
\be\widehat f_{_{\!\rm D}}^{\,\mu}=-\overline\nabla_{\!\nu}
\widehat T_{_{\!\rm D}}{^{\mu\nu}}\, ,\eqn{79}\fe
in terms of a regularised local dilatonic stress energy tensor given by
\be\widehat T_{_{\!\rm D}}{^{\mu\nu}}=\widehat\phi\big(\overline T{^{\mu\nu}}
-{_1\over ^4}\overline T\eta^{\mu\nu}+\overline{\cal C}{^{\mu\nu}}\big)
\, .\eqn{80}\fe
More specificly, in the particular case of a Nambu Goto type string, as
characterised by (\ref{62}) and (\ref{65}) it can be seen that this dilatonic
stress energy contribution will reduce to the form
\be\widehat T_{_{\!\rm D}}{^{\mu\nu}} =2\,\widehat l\
m_{_{\rm K}}^{\, 4}m_{_{\rm D}}^{-2}\eta^{\mu\nu}\, .\eqn{81}\fe

Comparing (\ref{81}) with (\ref{78}) it can be seen that (in contradiction
with previous assertions to the effect that it would vanish~\cite{CHH90} or
even that it would augment the corresponding axionic contribution
~\cite{DH89}) this dilatonic contribution must always be oppositely directed
to the corresponding axionic contribution. More particularly it can be seen
that these dominant local axionic and dilatonic self interaction contributions
will exactly cancel each other out if the relevant coupling constants are
related by the condition
\be 2 m_{_{\rm K}}^{\,2} =\overline\kappa m_{_{\rm D}}m_{_{\rm J}}
\, ,\eqn{82}\fe 
which will in fact be satisfied automatically in the special case (\ref{61})
envisaged by Dabholkar and Harvey~\cite{DH89} (whose faulty analysis lead to a
condition that was somewhat different from (\ref{82}) -- but that also
happenned to be satisfied in the special case (\ref{61}) they were
considering, and that thereby provided a spurious verification of their
reasonning.)

The mutual cancellation subject to (\ref{82}) of the dominant axionic and
dilatonic self interactions for a Nambu-Goto string in a 4-dimensional
background has already been confirmed by Buonanno and Damour~\cite{BD98a}
using an entirely different approach formulated in terms of an effective
action. The fact that -- as in the previously investigated
electromagnetic~\cite{C97} and gravitational~\cite{CB98,C98b} cases -- the
result of the present approach based on direct evaluation of the self force is
in full agreement with the result of the approach based on the use of an
effective action provides a check on the validity of the latter approach,
whose credibility had previously been questionned~\cite{CHH90}. The complete
consistency between the two kinds of approach has already been made clear for
the electromagnetic~\cite{C97} and gravitational~\cite{C98b} cases, and will
be made clear for the axionic and dilatonic cases dilatonic in the next
section, where the relevant effective action contributions will be explicitly
derived.

\section{Action renormalisation}

The fact that the dominant local force contributions are expressible as
divergences of the form (\ref{76}) and (\ref{79})  is what makes it possible
to describe the the result of this regularisation as a ``renormalisation'': it
implies that these self force contributions   can be absorbed into the left
hand side of the basic force balance equation by a renormalisation whereby the
original ``bare'' stress momentum energy density tensor $\overline
T{^{\mu\nu}}$ undergoes a replacement $\overline T{^{\mu\nu}}\mapsto\widetilde
T{^{\mu\nu}}$ in which the ``dressed'' stress momentum energy tensor is given
by
\be \widetilde T{^{\mu\nu}}=\overline T{^{\mu\nu}}+  \widehat T{^{\mu\nu}}
\eqn{83}\fe
with
\be  \widehat T{^{\mu\nu}}   =  \widehat T_{_{\!\rm J}}{^{\mu\nu}}
+ \widehat T_{_{\!\rm D}}{^{\mu\nu}}\, . \eqn{84}\fe
The basic force balance equation (\ref{51}) can thereby be rewritten in the
equivalent form
\be\overline\nabla_{\!\nu}\widetilde T{^{\mu\nu}}= \widetilde f {^\mu}
\eqn{85}\, ,\fe
in which the force density on the right consists just of
well behaved long range radiation contributions as given by the sum
\be \widetilde f {^\mu}=\widetilde f_{_{\!\rm J}}{^\mu}+
\widetilde f_{_{\!\rm D}}{^\mu}\, ,\eqn{86}\fe
in which each of the terms is entirely regular.

What will be shown in this final section is that the renormalised stress
energy tensor $\widetilde T{^{\mu\nu}}$ is derivable, by a prescription of the
standard form (\ref{19}), from a correspondingly remormalised action
 in which the original Lagrangian master
function $\overline{\cal L}_{_{\rm ma}}$ is replaced by an appropriately
renormalised master function $\widetilde{\cal L}_{_{\rm ma}}$.

In order to incorporate the effects of self interaction, as described by
the renormalised force balance equation (\ref{86}),
it can be verified that all one needs to do is to replace ${\cal I}_{_{\rm 
ma}}$ by a corresponding renormalised action
\be \widetilde{\cal I}_{_{\rm ma}}=\int\widetilde{\cal L}_{_{\rm ma}}\, 
\Vert\gamma\Vert^{1/2}\, d^2\sigma \, ,\eqn{88}\fe
with
\be\widetilde{\cal L}_{_{\rm ma}}=
\widetilde{\cal L}
+{_1\over^2} \widetilde B_{\mu\nu}\overline W{^{\mu\nu}}
+\widetilde \phi\overline T  \, .\eqn{89}\fe
where the renormalised master function is given simply by
\be\widetilde{\cal L}={\cal L}+\widehat\Lambda_{_{\rm J}}+
\widehat{\cal L}_{_{\rm D}}  \, , \eqn{90}\fe
in which the  axionic contribution will be given by
\be\widehat{\cal L}_{_{\rm J}}={_1\over^4}\widehat B_{\mu\nu}
\overline W{^{\mu\nu}}\ ,
\eqn{91} \fe
which works out simply to be a constant,
\be\widehat{\cal L}_{_{\rm J}}=-{_1\over ^2} \widehat l\ \bar\kappa^2
m_{_{\rm Z }}^{\, 2}\, .\eqn{92}\fe
in which it is to be recalled that $\widehat l$ is the logarithmic
factor given by (\ref{70}).
The corresponding dilatonic contribution (which has not been evaluated
for a general string model before) is obtained as
\be \widehat{\cal L}_{_{\rm D}}={_1\over^2}\widehat \phi\,
\overline T ={_1\over^2} \widehat l\
m_{_{\rm D}}^{-2}\,
\overline T^2 \, .\eqn{93}\fe

It is to be observed that in terms of the string energy density
${\cal U}$ and tension ${\cal T}$ (as conventionaly defined to
be the eigenvalues of $-\overline T{^{\mu\nu}}$) the trace in the proceeding
formula will be given by $\overline T=-({\cal U}+{\cal T})$,
so the dilatonic self interaction contribution will be proportional
to the $({\cal U}+{\cal T})^2$.  This is to be contrasted with the case
of the corresponding gravitational self interaction contribution
which has been shown~\cite{C98a} to be proportional to 
$({\cal U}-{\cal T})^2$. For a Nambu-Goto string, as characterised by
${\cal U}={\cal T}=m_{_{\rm K}}^{\,2}$ this gravitational contribution
will simply vanish, while the dilatonic contribution (\ref{93}) will just
have the constant value given by
\be \widehat{\cal L}_{_{\rm D}}= 2 \widehat l\ m_{_{\rm K}}^{\,4}
m_{_{\rm D}}^{-2}\, \, .\eqn{94}\fe
The self interaction contributions (\ref{92}) and (\ref{94}) are in perfect
agreement with those already obtained by the effective action approach
developed by Buonanno and Damour~\cite{BD98a} -- whose results were more
general than those provided here in so much as they were not limited to a
4-dimensional background, though on the other hand they were more restricted
in so much as they considered only strings of Nambu-Goto type.

This completes the demonstration that, as has already been shown
for electromagnetic~\cite{C97} and gravitational~\cite{C98b} interactions
(and contrary to what was previously alleged~\cite{CHH90}) so also for
the axionic and dilatonic contributions, the treatment of the divergent self 
interaction contribution by an approach developped directly in terms
of effective action is entirely consistent with the treatment based
on the detailed analysis of the corresponding force contributions,
as given in the axionic case by (\ref{66}) and in the dilatonic case
by the general formula (\ref{54}) or its Nambu-Goto specialisation
(\ref{67}). 

The discrepancies that arose in earlier work were due to the use of unreliable
short cut methods (largely motivated by the unavailability of the more
efficient methods of geometrical analysis that have since been developed)
which lead to the omission of some of the (less easily calculable) terms in
the relevant formulae. In particular the sign error in the original estimate
of the net dilatonic contribution by Dabholkar and Harvey~\cite{DH89} can be
accounted for as being due to omission of the contribution provided by the
first term on the right in (\ref{67}). As was seen for the corresponding
forces in the previous section, the axionic contribution (\ref{92}) will
evidently be exactly cancelled by the dilatonic contribution (\ref{94}) when
the special condition (\ref{82}) is satisfied. Like the self cancellation of
the gravitational contribution in the Nambu=Goto case~\cite{CB98,C98b}, this
mutual cancellation of the corresponding axionic and dilatonic contributions
is a special feature of 4-dimensional space-time: it has been shown by the
work of Buonanno and Damour~\cite{BD98a} that it does not carry over to
Nambu-Goto strings in backgrounds of higher dimension.

I wish to thank A. Buonanno, T. Damour, R. Battye, A. Gangui,
G. Gibbons, P. Peter and P. Shellard for instructive conversations.

\end{document}